# LASER STABILIZATION CONTROLS FOR THE ISAC BEAM POLARIZER

R. Nussbaumer, D. Dale, H. Hui, J. Richards, and P. Levy,
TRIUMF, Vancouver, BC, V6T 2A3, Canada

Abstract

The ISAC polarized beam facility uses a Ti:Sapphire laser for producing spin-polarized beams of short-lived radioactive isotopes, initially 7Li. The laser power and spectral content must be tightly controlled and monitored. The control system to accomplish this task uses techniques that combine operator interface, data reduction, and closed loop automation. The laser stabilization application uses input data acquired from a spectrum analyzer, in the form of a waveform array repetitively acquired from a GPIB-interfaced oscilloscope. A Labview based operator interface allows the selection of features from within the waveform, which are then used as the basis for closed loop control, in concert with the existing EPICS control system software and related hardware. Part of the challenge stems from using input data which are subject to human interpretation and require distillation into a hard numeric format suitable for use in a control system.

## 1 PREAMBLE

The ISAC radioactive beam facility at TRIUMF produces spin-polarized beams short-lived exotic ions. An Argon laser is used in an optical pumping arrangement, exciting a Ti:Sapphire laser. The stability of the Ti:Sapphire laser is insufficient to achieve the degree of polarization needed by experiments, for periods of more than a few seconds. An external control system has been devised and implemented to enhance the stability of the laser.

The laser fundamental frequency is controlled by a piezo-electric element, which adjusts the laser cavity length. An etalon optical bandpass filter controls the laser output frequency. The etalon passband is controlled by changing the angle of incidence of the laser light, using a servo motor. These two parts of the laser are under control of the stabilization system.

The laser spectral content is measured by a spectrum analyzer which is subject to a slow measurement drift. Helium:Neon laser light is mixed into the Ti:Sapphire laser light as a stable reference, to measure the drift. Modulating the spectrum analyzer bias input compensates for this slow drift.

The spectrum analyzer output is digitized by a Digital Storage Oscilloscope and read via an ethernet-attached GPIB interface. Figure 1 illustrates the arrangement of the principal components of the system.

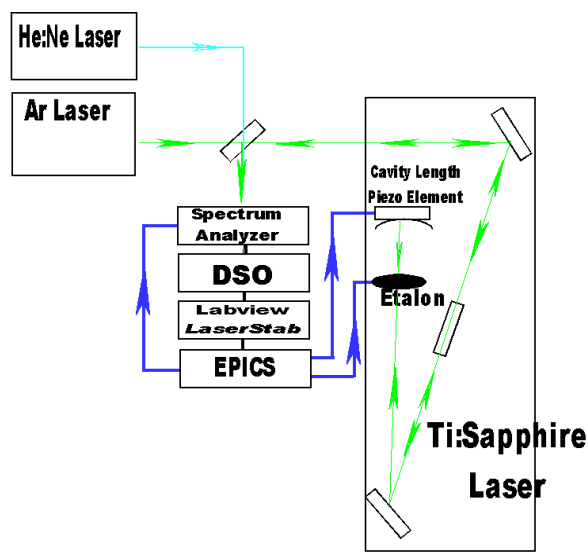

Figure 1: Block diagram of the Laser Stabilization Control System

## 2 DATA

One of the principal challenges in the development of the laser stabilization system is reliably extracting the features used to control the laser.

### 2.1 Jitter

Jitter within the spectral content of the Ti:Sapphire laser appears as frequency shifts of 15-25 X-axis units between sequential readings from the DSO. The primary source of this jitter is acoustical noise coupled into the laser table.

### 2.2 Imprecise Data

The general shape of the waveform features is not necessarily clear or distinct in nature. The baseline value of the waveform is arbitrarily set by operators in the initial setup, is subject to drift, and contains noise. Moreover, features within a single scan of the spectrum analyzer must be compared to previous corresponding features, in order to determine what changes are slowly occurring. This differs from typical control systems, where individual transducers produce distinct signals, the transducer is read iteratively, and no further extraction of the signal (other than scaling, or perhaps noise filtering) is required.

## 2 SOFTWARE

The software component of the laser stabilization system is integrated into the EPICS based ISAC control system and consists of a combination of EPICS running on VME IOCs, and a Sun Solaris based Labview application. Labview based software was written to provide an interactive operator interface. Labview was chosen primarily for its streamlined access to the GPIB interfaced DSO, and because it was already available at the onset of the project. National Instruments provides a clean API for accessing GPIB based instruments, that is accessible from Labview and from conventional high level languages such as C.

Some EPICS Display Manager (DM) operator interface software is used, mainly for passive monitoring of the laser waveform and control signals.

## 3 WAVEFORM AQUISITION AND ANALYSIS

Waveforms are acquired as 10-sample bursts of 500 integer data point waveforms from the DSO, as unit-less data.

### 3.1 Simple Filtering

An initial strategy to cope with the frequency jitter in the Ti:Sapphire laser was to allow the Tektronix DSO to perform some averaging. While being simple to deploy, this approach was found to be ineffective. The DSO averages sample points that are correlated in time, resulting in waveshapes that appear to contain multiple peaks. However, the intent was to obtain a single average X-axis position of the power peaks within the frequency spectrum.

More appropriate averaging is performed in software by acquiring multiple discrete waveforms sampled as quickly as possible. Each of the waveforms is then discretely analyzed, and the positions and amplitudes of peaks within the waveforms are extracted.

### 3.2 Peak detection

Analysis of the individual waveforms principally entails peak detection. A C language subroutine was written to accomplish this task, taking peculiarities of the application into consideration. The algorithm efficiently distinguishes even small peaks from a slowly varying background by using a differential method. This improves on the built-in Labview peak detection function, which requires a data-dependent threshold value at runtime.

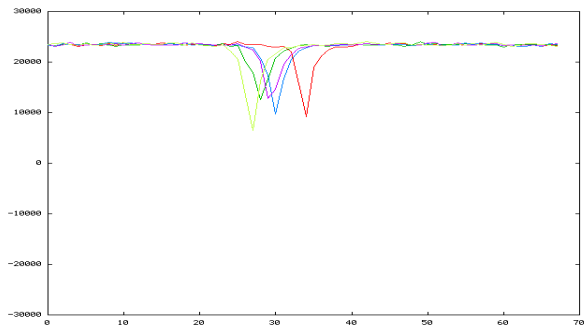

Figure 2: Five waveforms sampled in close succession, showing jitter in peak position along the X (frequency) axis

### 3.3 Peak Averaging

Ten waveforms are sampled and peaks are found within each waveform. The individual peaks within each waveform are correlated with equivalent peaks in all others, yielding a single set of waveform peak positions and the corresponding amplitudes. Figure 2 shows the relative spacing of the peaks for five waveforms sampled in fast succession.

### 3.4 Peak Tracking

In order to control the various stabilizing elements in the system, peaks are tracked and their x-axis movements are measured. Because there may be differing numbers of peaks from one overall scan to the next, it is necessary to correlate peaks in consecutive waveforms with the corresponding peaks in the previous scan. As new peaks emerge over time, they must be detected as such.

Peaks that exist consistently over time must be selectable, either automatically, or by a human operator. Selected peaks are used as the control points in the stabilization system.

## 4 CONTROL STRATEGY

### 4.1 Spectrum Analyzer

The spectrum analyzer bias is controlled by monitoring the He:Ne reference peak position, with respect to the position at the start time of the closed loop control. The bias voltage is adjusted proportional to the X-axis drift. This control action is performed in an EPICS subroutine record.

### 4.2 Cavity Length Piezo Element

The laser cavity length stabilization is performed by monitoring the X-axis distance between the He:Ne reference peak and the one peak selected as the cavity length control peak. When the distance differs from the distance at the start of closed-loop control mode, a corrective signal proportional to the drift is applied. This control action is performed in an EPICS subroutine record.

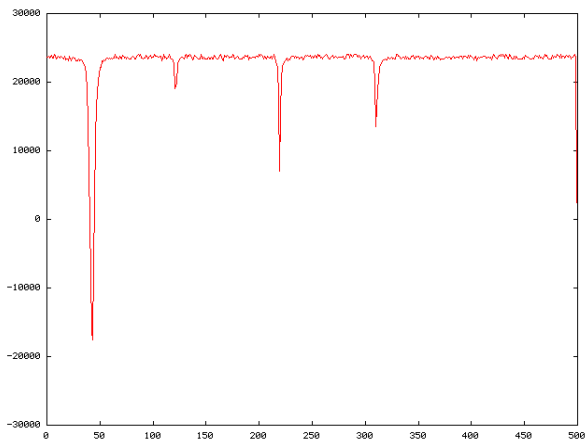

Figure 3: A typical waveform showing the emergence of a third Ti:Sapphire peak, to be suppressed by modulating the etalon  The He:Ne reference peak is at the left of the graph.

### 4.3 Etalon

The more complex etalon control strategy is modeled after the actions of a human operator. The etalon, as a bandpass filter, allows two or more bands of laser power to pass.  When the etalon is optimally positioned, the two Ti:Sapphire peaks are about equal in magnitude, and there are no side peaks.  The human operator notices the emergence of side peaks and adjusts the control voltage appropriately. The software similarly anticipates the location of the side peaks, and when peaks are detected, a suppressive signal, proportional to the amplitude of the side peak., is applied.

Spurious noise spikes are discriminated against, by anticipating the location of the side peaks and ignoring any that appear but are not within the expected range along the X-axis. When the anticipated location of side peaks would overlap the He:Ne reference peak, a warning message is displayed, to prompt the operator to make some manual adjustment.

## 5 EVALUATION

The system has been successfully used to stabilize the laser without user intervention for durations of several hours, and has not been a limiting factor in the duration of polarized beam production.  A few deficiencies have, however, been identified.

The use of high level software like Labview is not optimal for an application that must perform in a real time control application. High sample rates cannot be achieved, and the non-real time host operating system may introduce unacceptable delays.  There have been problems with the reliability of the GPIB-Enet hardware when operating for the extended periods required in this application.  It is not understood from where this problem originates.

The existing combination of the Labview *LaserStab* application and EPICS hosted subroutine records is a sub-optimal arrangement from the perspective of software maintenance.  Of particular concern is the current architecture, which does not allow for multiple instances of the Labview application to run concurrently on multiple workstations. Labview programming is not well understood by other TRIUMF staff, and is not a good fit with existing tools, such as version control systems.

## 6 SUMMARY

The low frequency drift of the Ti:Sapphire laser and its measurement system can be successfully controlled, using primitive software control algorithms.  With proper consideration for anomalous conditions, waveform data can be adequately distilled into discrete reference points for closed loop control.

The next step in the evolution of the system will increase the overall bandwidth of the measurement and control system to attempt to eliminate higher frequency noise. This will require the use of a robust real time operating system to achieve faster and more accurate sampling and control.